\documentclass[twocolumn]{aastex631}

\usepackage{longtable}
\usepackage{threeparttablex}
\usepackage{booktabs}
\usepackage{tablefootnote}

\title{HI observations of 100 nearby dwarfs}
\shortauthors{Nazarova et al.}

\begin{document}

\title{100 Proof:  HI Observations of 100 Nearby Dwarf Galaxies with the 100-meter Green Bank Telescope}


\correspondingauthor{Aleksandra E. Nazarova}
\email{a.e.nazarova@sao.ru}

\author{Aleksandra E.\ Nazarova}
\affiliation{Special Astrophysical Observatory, The Russian Academy of Sciences, Nizhnij Arkhyz, Karachai-Cherkessian Republic 369167, Russia}

\author{John M.\ Cannon}
\affiliation{Physics and Astronomy Department, Macalester College, 1600 Grand Avenue, Saint Paul, MN 55105, USA}
 
\author{Igor D.\ Karachentsev}
\affiliation{Special Astrophysical Observatory, The Russian Academy of Sciences, Nizhnij Arkhyz, Karachai-Cherkessian Republic 369167, Russia}

\author{Dmitry I.\ Makarov}
\affiliation{Special Astrophysical Observatory, The Russian Academy of Sciences, Nizhnij Arkhyz, Karachai-Cherkessian Republic 369167, Russia}
 
\author{Maksim I.\ Chazov}
\affiliation{Special Astrophysical Observatory, The Russian Academy of Sciences, Nizhnij Arkhyz, Karachai-Cherkessian Republic 369167, Russia}

\author{Lila Schisgal}
\affiliation{Physics and Astronomy Department, Macalester College, 1600 Grand Avenue, Saint Paul, MN 55105, USA}

\author{William St.\ John}
\affiliation{Physics and Astronomy Department, Macalester College, 1600 Grand Avenue, Saint Paul, MN 55105, USA}

\begin{abstract}

We describe the results of observations with the 100 m Robert C. Byrd Green Bank Telescope (GBT) in the HI line of 105 nearby dwarf galaxies, 60 of which were discovered recently in the DESI Legacy Imaging Surveys. Of 105 objects observed, we detected 77 galaxies with the following median parameters: an HI-flux of 0.69~Jy\,km\,s$^{-1}$, a heliocentric velocity of 732~km\,s$^{-1}$, and a $W_{50}$ line width of 32~km\,s$^{-1}$. 70 are isolated late-type objects and 35 are new probable satellites of nearby spiral galaxies (NGC\,628, NGC\,2787, NGC\,3556, NGC\,4490, NGC\,4594 and NGC\,5055). The detected galaxies are  predominantly gas-rich systems with a median gas-to-stellar-mass ratio of 1.87. In general, they follow the classic Tully-Fisher relation obtained for large disk-dominated spiral galaxies if their $M_{21}$ magnitudes are used instead of B-magnitudes.

\end{abstract}


\section{Introduction}

Most existing galaxy catalogs are samples, limited by an apparent magnitude or flux of objects in a fixed spectral range. However, the results of modeling the large-scale structure of the Universe within the framework of the standard cosmological model $\Lambda$CDM need to be compared with an ensemble of galaxies limited to a fixed volume. The most suitable sample of this type is the Updated Nearby Galaxy Catalog~\citep[UNGC,][]{kar2013}, a regularly updated version\footnote{\url{http://www.sao.ru/lv/lvgdb}} of which contains about 1500 galaxies with distances $D < 12$~Mpc.  
It is obvious that the population of this Local Volume (LV) sample can either increase with the advent of deeper sky surveys, or decrease somewhat as the distances of galaxies are refined. The radius of the LV sphere, $\sim$12~Mpc, is defined by the ability to  measure the distance of an individual galaxy with an error of 5\% using the Tip of the Red Giant Branch~\citep[TRGB,][]{1993ApJ...417..553L, 10.1093/mnras/staa3668} method in a single-orbit observation with the Hubble Space Telescope (HST). A great advantage of the UNGC catalog is its predominance of low-luminosity dwarf galaxies, which are usually inaccessible to observations at larger distances. The wealth of observational data on the radial velocities and distances of LV galaxies provides the opportunity to study the local field of peculiar velocities, which is determined by the position and masses of local attractors (i.e., it traces the dark matter distribution on small cosmic scales).

An effective way to gather information about the radial velocities of LV galaxies is via ``blind'' surveys of large areas of the sky in the 21-cm hydrogen line, such as HIPASS~\citep{kor2004} and ALFALFA~\citep{hay2018}. These surveys predominantly studied regions with declination $\mathrm{Dec}< +38^{\circ}$. More northerly coverage is provided by the WSRT HI survey of the Canes Venatici region \citep{kov2009}, by the FAST radio telescope~\citep{jia2020,zha2024} and by the Apertif Shallow HI Survey~\citep{2024A&A...692A.217S, 2022A&A...667A..38A}.

An additional opportunity to detect dwarf galaxies in the LV appeared with the publication of DESI Legacy Imaging Surveys~\citep{dey2019}. The search for new LV objects~\citep{kar2022,kara2023,kar2024} led to the discovery of more than a hundred dwarf galaxies at high declination. About half of them are dwarf spheroidal systems (dSph)  with a low content of neutral hydrogen. After excluding dSphs, we selected 54 dwarf galaxies with evidence of active star formation to measure their radial velocities with the Green Bank Telescope (GBT). We also added to this sample 51 galaxies from the LV that had no HI-parameter estimates. During the implementation of our program, new HI data from the FASHI survey appeared~\citep{zha2024, 2024A&A...684L..24K}. Comparison of these independent data  makes it possible to estimate the measurement accuracy of the radial velocity and other parameters of the observed galaxies.

Our article is organized as follows. Section~\ref{sec:GBT} contains a description of observations at the GBT. Section~\ref{sec:HIresults} presents the results of observations of our targets in the HI line. Section~\ref{sec:Notes} briefly describes the individual features of the observed galaxies and their environment. Section~\ref{sec:Discussion} contains a discussion of the results obtained in the context of the characteristics  of other LV late-type dwarf galaxies.  Conclusions are presented in Section~\ref{sec:Conclusion}.

\ 

\section{GBT observations of the nearby dwarfs.}
\label{sec:GBT}

We observed 105 dwarf galaxies with the National Radio Astronomy Observatory 100m Robert C.\ Byrd Green Bank Telescope (GBT\footnote{The National Radio Astronomy Observatory and Green Bank Observatory are facilities of the U.S. National Science Foundation operated under cooperative agreement by Associated Universities, Inc.}) under projects GBT/23B--032 (P.I.\ Cannon) and GBT/24B-157 (P.I.\ Nazarova) with the VEGAS (Versatile GBT Astronomical Spectrometer) backend. ON-OFF position switching observations were performed for all sources; 32,768 channels covered a 23.44 MHz bandpass that was centered at 1410.93 MHz (the rest frequency of HI redshifted by 2000~km\,s$^{-1}$). The total effective frequency coverage was from $\sim1399.2$ MHz to $\sim1422.6$~MHz. This covers the approximate velocity range from $-470$ to $+4470$~km\,s$^{-1}$, which is adequate to probe the LV within 10~Mpc. RFI was minimal in the frequency range of interest. Velocities are presented in the optical convention $cz$. Data were acquired during 30 observing sessions in the second half of 2023 and 2024 (100 hours in total). The time per target was typically 40 minutes, 20 minutes of which was spent at the ON-position. The duration could vary markedly depending on the HI brightness.

All reductions were performed in the IDL environment\footnote{Exelis Visual Information Solutions, Boulder, CO.}, using the \textsc{GBTIDL} package designed at NRAO. For most galaxies, velocity and width measurements, along with their errors, were obtained by fitting a single gaussian to the line profile using the \textit{fitgauss} procedure after fitting and subtracting the local continuum. The resulting spectra have an median rms noise of 3.3 mJy after smoothing to a velocity resolution of 1.81~km~s$^{-1}$ using a \textit{boxcar} function. The total flux in the line is calculated as the area under the profile using the \textit{stats} procedure. The corresponding flux error was estimated as the total variance of the background fluctuations. This is true in the case of low flux integrals compared to the background, which is realized in our observations. In two cases, Dw\,1247$-$0824 and Dw\,1459+44, there was a double horn line and measurements were taken using the \textit{gmeasure} procedure. For KUG\,1033+414 this \textit{gmeasure}-approach resulted in large measurement errors, so the entry in Table~\ref{table1} contains data for a single gaussian fit.

\section{Observational results.}
\label{sec:HIresults}
       
\begin{figure*}[!ht]
\centering
\caption{Images of 77 HI-detected nearby late-type dwarf galaxies taken from DESI Legacy Imaging Surveys. Each image side is $2\arcmin$, North is to the top and East is to the left.}
\includegraphics[width=0.8\textwidth]{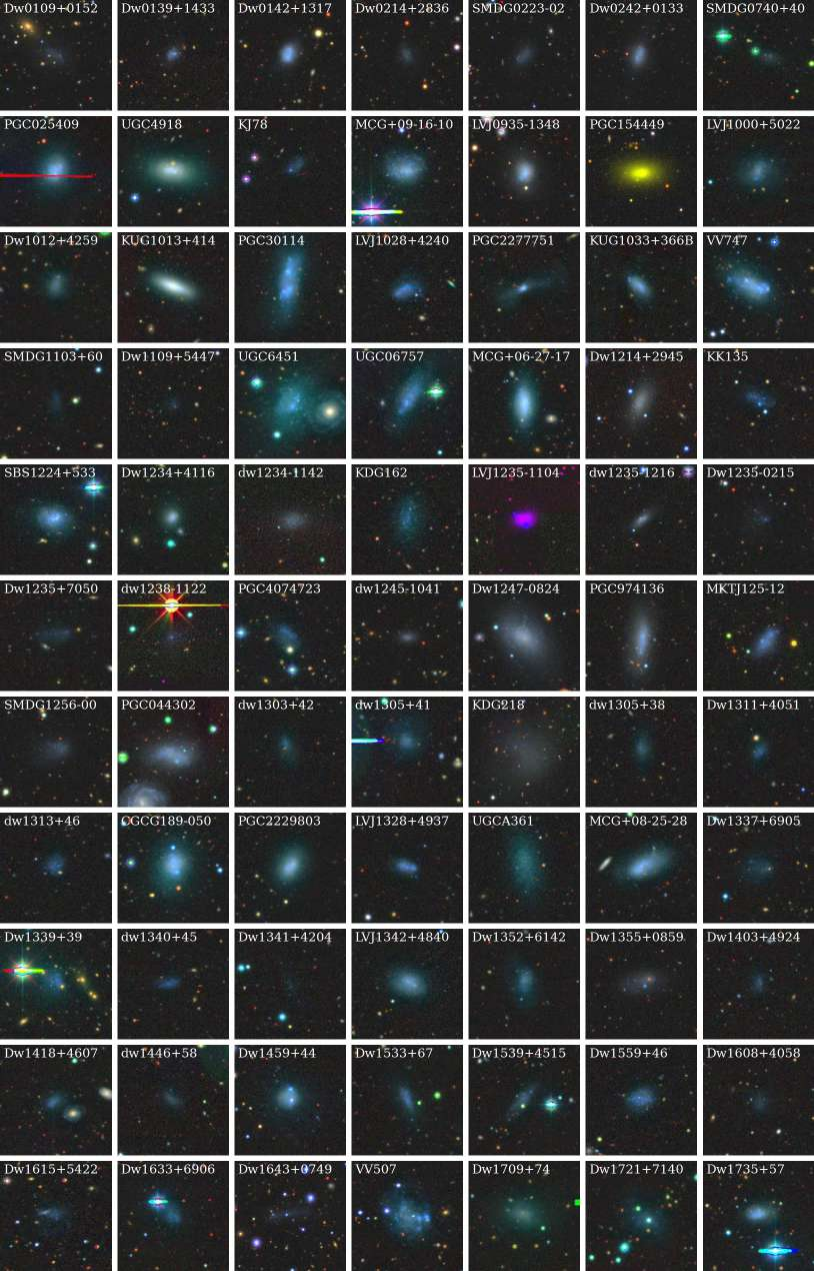}
\label{fig:mosaic77}
\end{figure*}

Of the 105 GBT targets, the HI signal was detected in 77 dwarf galaxies.
Their list is presented in Table~\ref{table1}, whose columns contain the following: 
(1) the name of the galaxy as indicated in the UNGC~\citep{kar2013};
(2) equatorial  coordinates for the epoch J2000.0;
(3) apparent optical axial ratio of the galaxy;
(4,5) thr integral apparent magnitudes in the $B$ and $V$ bands estimated from $g$ magnitude and $(g - r)$ color, which were measured by us using the DESI Legacy Imaging Surveys\footnote{\url{https://www.legacysurvey.org/}}~\citep{dey2019} images and the transformation equations from The Dark Energy Survey~\citep[DESI,][]{abb2021}, and in several cases $B$ magnitudes were taken from the database HyperLEDA~\citep{mak2014}; 
(6) heliocentric radial velocity of the galaxy (km s$^{-1}$) with indication of its error;  
(7) full width half maximum of HI line (km\,s$^{-1}$) with indication of its error;
(8) integral flux in the HI line (Jy\,km\,s$^{-1}$) with indication of its error; 
(9) kinematic distance of the galaxy in Mpc, calculated via its radial velocity with account of local streams according to the Numerical Action Method model~\citep[NAM,][]{sha2017,kou2020}; 
(10,11) hydrogen and stellar mass (in solar masses) determined in Section~\ref{sec:Discussion}.

\begin{figure*}
\centering
\caption{HI spectra of 77 dwarf galaxies detected with GBT. The HI flux integrals and $W_{50}$ values are given above each panel.} 
\includegraphics[width=1\linewidth]{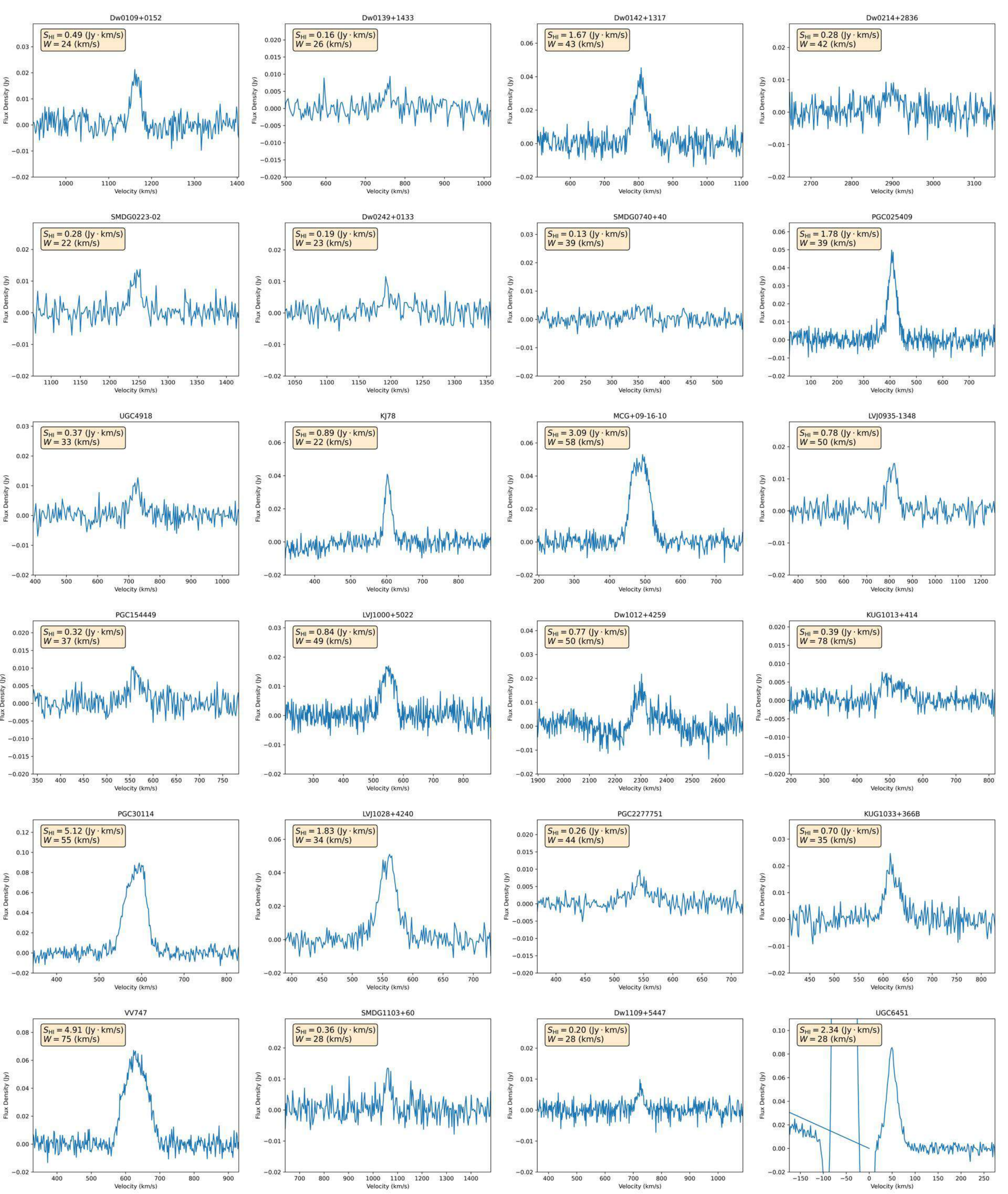}
\label{fig:spectra}
\end{figure*}

The $2\arcmin\times2\arcmin$ images (North up and East to the left) from the DESI Legacy Imaging Surveys of the 77 detected sources are presented in Fig.~\ref{fig:mosaic77}. 
Their spectra are shown in Fig.~\ref{fig:spectra}. Insets in each panel provide S$_{\rm HI}$ and W$_{\rm 50}$ values.

\setlength{\tabcolsep}{3.8pt}
\begin{ThreePartTable}
\begin{TableNotes}
\item[ $ $ ] \textbf{Bold font} indicates that the galaxy is within the LV ($D_\mathrm{NAM} < 12 $ Mpc).
\end{TableNotes}

\begin{longtable*}{l c c c c r c c c c c} 
\caption{List of galaxies detected with GBT.\label{table1}} \\
\insertTableNotes \\
\hline\hline
\multicolumn{1}{c}{Name}            &  
RA(2000.0) Dec.                     &   
$b/a$                               &    
$B$                                 &  
$B - V$                             & 
\multicolumn{1}{c}{$V_\mathrm{h}$}  & 
\multicolumn{1}{c}{$W_{50}$}        & 
\multicolumn{1}{c}{$S_\mathrm{HI}$} & 
$D_\mathrm{NAM}$                    & 
$\log{M_{\rm HI}}$                  & 
$\log{M_*}$                         \\

                                    &
                                    &
                                    &
\multicolumn{1}{c}{mag}             &
\multicolumn{1}{c}{mag}             &
\multicolumn{1}{c}{km\,s$^{-1}$}    &
\multicolumn{1}{c}{km\,s$^{-1}$}    &
\multicolumn{1}{c}{Jy\,km s$^{-1}$} &
\multicolumn{1}{c}{Mpc}             &
\multicolumn{1}{c}{$M_\sun$}        &
\multicolumn{1}{c}{$M_\sun$}        \\
\hline
\endfirsthead

\caption{continue}\\
\hline
\multicolumn{1}{c}{Name}            &  
RA(2000.0) Dec.                     &   
$b/a$                               &    
$B$                                 &  
$B - V$                             & 
\multicolumn{1}{c}{$V_\mathrm{h}$}  & 
\multicolumn{1}{c}{$W_{50}$}        & 
\multicolumn{1}{c}{$S_\mathrm{HI}$} & 
$D_\mathrm{NAM}$                    & 
$\log{M_{\rm HI}}$                  & 
$\log{M_*}$                         \\

                                    &
                                    &
                                    &
\multicolumn{1}{c}{mag}             &
\multicolumn{1}{c}{mag}             &
\multicolumn{1}{c}{km s$^{-1}$}     &
\multicolumn{1}{c}{km s$^{-1}$}     &
\multicolumn{1}{c}{Jy km s$^{-1}$}  &
\multicolumn{1}{c}{Mpc}             &
\multicolumn{1}{c}{M$\sun$}         &
\multicolumn{1}{c}{M$\sun$}         \\
\hline
\endhead

\hline
\endfoot

\hline\hline
\endlastfoot

Dw\,0109+0152	              & 01:09:17.0+01:52:59	  &	0.54 &	18.24	&	0.61	&	1163.5$\pm$0.7	&	24.0$\pm$1.5	&	0.49$\pm$0.04	&	15.0	&	7.41	&	7.21	\\
\textbf{Dw\,0139+1433}	  & 01:39:50.6+14:33:22	  &	0.87 &	18.25	&	0.55	&	 757.1$\pm$2.8	&	26.2$\pm$6.6	&	0.16$\pm$0.04	&	10.8	&	6.64	&	6.84	\\
\textbf{Dw\,0142+1317}	  & 01:42:07.2+13:17:38	  &	0.83 &	16.72	&	0.48	&	 804.0$\pm$0.8	&	43.2$\pm$1.9	&	1.67$\pm$0.07	&	11.1	&	7.69	&	7.38	\\
Dw\,0214+2836	              & 02:14:09.6+28:36:47	  &	0.48 &	18.98	&	0.62	&	2899.1$\pm$2.6	&	41.5$\pm$6.2	&	0.26$\pm$0.04	&	36.1	&	7.90	&	7.70	\\
SMDG\,0223$-$02	          & 02:23:18.7$-$02:03:25 &	0.67 &	18.61	&	0.57	&	1245.1$\pm$0.9	&	21.9$\pm$2.2	&	0.28$\pm$0.03	&	14.7	&	7.15	&	6.99	\\
Dw\,0242+0133	              & 02:42:33.1+01:33:50	  &	0.73 &	17.37	&	0.60	&	1195.6$\pm$1.7	&	23.0$\pm$4.0	&	0.19$\pm$0.03	&	14.7	&	6.98	&	7.53	\\
\textbf{SMDG\,0740+40}	  & 07:40:23.0+40:32:56	  &	0.63 &	19.18	&	0.81	&	 353.5$\pm$3.6	&	38.8$\pm$8.6	&	0.13$\pm$0.02	&	10.1	&	6.49	&	6.79	\\
\textbf{PGC\,025409}	  & 09:02:50.6+71:18:22	  &	0.73 &	15.79	&	0.48	&	 408.4$\pm$0.5	&	38.6$\pm$1.2	&	1.81$\pm$0.05	&	10.3	&	7.65	&	7.69	\\
UGC\,4918	              & 09:19:17.7+69:48:04	  &	0.93 &	15.73	&	0.85	&	 723.3$\pm$1.5	&	32.9$\pm$3.6	&	0.36$\pm$0.03	&	14.0	&	7.22	&	8.51	\\
KJ78	                  & 09:20:36.4+49:40:31	  &	0.62 &	18.42	&	0.44	&	 603.8$\pm$0.4	&	22.0$\pm$0.9	&	0.89$\pm$0.04	&	13.2	&	7.56	&	6.78	\\
\textbf{MCG+09-16-10}	  & 09:23:17.0+51:58:22	  &	0.78 &	16.09	&	0.58	&	 485.5$\pm$0.7	&	57.5$\pm$1.5	&	3.09$\pm$0.06	&	11.0	&	7.94	&	7.76	\\
\textbf{LV\,J0935$-$1348} & 09:35:21.6$-$13:48:52 &	0.75 &	16.44	&	0.67	&	 811.3$\pm$1.7	&	50.4$\pm$4.1	&	0.78$\pm$0.06	&	11.7	&	7.40	&	7.81	\\
\textbf{PGC\,154449}	  & 09:57:08.9$-$09:15:48 &	0.93 &	15.98	&	0.86	&	 562.7$\pm$1.8	&	36.6$\pm$4.2	&	0.32$\pm$0.03	&	 8.0	&	6.68	&	7.94	\\
\textbf{LV\,J1000+5022}	  & 10:00:25.5+50:22:45	  &	0.76 &	16.99	&	0.66	&	 548.7$\pm$1.2	&	48.7$\pm$2.8	&	0.84$\pm$0.05	&	11.9	&	7.45	&	7.58	\\
Dw\,1012+4259	              & 10:12:42.7+42:59:31	  &	0.75 &	17.87	&	0.72	&	2297.0$\pm$1.7	&	49.5$\pm$4.2	&	0.74$\pm$0.04	&	37.0	&	8.37	&	8.30	\\
\textbf{KUG\,1013+414}      & 10:16:15.6+41:09:58	  &	0.39 &	15.58	&	0.77	&	 506.6$\pm$3.2	&	78.4$\pm$7.5	&	0.39$\pm$0.03	&	 9.9	&	6.95	&	8.14	\\
\textbf{PGC\,30114}	      & 10:18:43.0+46:02:44	  &	0.30 &	15.21	&	0.42	&	 586.6$\pm$0.6	&	55.2$\pm$1.3	&	5.12$\pm$0.07	&	11.9	&	8.23	&	7.94	\\
\textbf{LV\,J1028+4240}   & 10:28:33.0+42:40:07	  &	0.71 &	17.00	&	0.36	&	 559.0$\pm$0.6	&	34.2$\pm$1.3	&	1.83$\pm$0.07	&	10.6	&	7.68	&	7.04	\\
\textbf{PGC\,2277751}	  & 10:35:12.1+46:14:12	  &	0.63 &	17.15	&	0.60	&	 544.0$\pm$1.9	&	43.9$\pm$4.5	&	0.26$\pm$0.03	&	10.8	&	6.85	&	7.35	\\
KUG\,1033+366B	          & 10:36:17.6+36:25:31	  &	0.53 &	17.05	&	0.58	&	 618.2$\pm$0.9	&	34.7$\pm$2.1	&	0.70$\pm$0.04	&	12.3	&	7.40	&	7.47	\\
\textbf{VV747}	      & 10:57:47.0+36:15:39	  &	0.56 &	15.49	&	0.46	&	 630.2$\pm$0.7	&	75.4$\pm$1.6	&	4.91$\pm$0.08	&	 9.8	&	8.05	&	7.73	\\
SMDG\,1103+60	          & 11:03:56.4+60:29:53	  &	0.64 &	19.24	&	0.46	&	1060.7$\pm$1.6	&	27.6$\pm$3.7	&	0.38$\pm$0.05	&	18.9	&	7.51	&	6.79	\\
Dw\,1109+5447	              & 11:09:13.2+54:47:10	  &	0.78 &	19.80	&	0.53	&	 724.4$\pm$1.0	&	27.8$\pm$2.4	&	0.20$\pm$0.02	&	13.9	&	6.95	&	6.41	\\
\textbf{UGC\,6451}	      & 11:28:46.4+79:36:07	  &	0.53 &	15.44	&	0.67	&	  49.6$\pm$0.3	&	28.0$\pm$0.7	&	2.36$\pm$0.03	&	 3.2	&	6.75	&	7.07	\\
\textbf{UGC\,06757}	      & 11:46:59.1+61:20:05	  &	0.55 &	16.09	&	0.50	&	  88.4$\pm$0.3	&	25.4$\pm$0.8	&	1.87$\pm$0.05	&	 2.6	&	6.48	&	6.39	\\
\textbf{MCG+06-27-17}	  & 12:09:56.4+36:26:07	  &	0.31 &	15.52	&	0.58	&	 326.8$\pm$0.8	&	49.3$\pm$1.9	&	1.94$\pm$0.07	&	 4.3	&	6.93	&	7.17	\\
\textbf{Dw\,1214+2945}	  & 12:14:26.6+29:45:50	  &	0.45 &	17.27	&	1.00	&	 457.5$\pm$2.7	&	33.4$\pm$6.4	&	0.19$\pm$0.03	&	 5.2	&	6.07	&	7.24	\\
\textbf{KK135}	      & 12:19:34.7+58:02:34   &	0.52 &	17.74	&	0.36	&	 142.5$\pm$0.5	&	20.4$\pm$1.1	&	0.70$\pm$0.04	&	 2.9	&	6.14	&	5.62	\\
\textbf{SBS\,1224+533}	  & 12:26:52.6+53:06:19	  &	0.69 &	16.00	&	0.46	&	 292.1$\pm$0.5	&	31.6$\pm$1.2	&	1.18$\pm$0.04	&	 6.2	&	7.03	&	7.12	\\
\textbf{Dw\,1234+4116}	  & 12:34:38.2+41:16:34	  &	0.81 &	17.28	&	0.76	&	 614.3$\pm$1.4	&	28.1$\pm$3.4	&	0.19$\pm$0.03	&	 8.9	&	6.53	&	7.36	\\
\textbf{dw\,1234$-$1142}	  & 12:34:48.6$-$11:42:25 &	0.84 &	17.61	&	0.69	&	1153.2$\pm$1.6	&	21.6$\pm$3.9	&	0.23$\pm$0.04	&	 9.5	&	6.68	&	7.19	\\
\textbf{KDG\,162}	      & 12:35:01.6+58:23:08	  &	0.69 &	17.28	&	0.63	&	 125.3$\pm$0.8	&	21.8$\pm$1.9	&	0.53$\pm$0.03	&	 2.9	&	6.01	&	6.18	\\
\textbf{LV\,J1235$-$1104} & 12:35:39.4$-$11:04:01 &	0.90 &	16.86	&	0.22	&	1099.7$\pm$0.9	&	55.4$\pm$2.1	&	1.97$\pm$0.08	&	 8.8	&	7.55	&	6.73	\\
\textbf{dw\,1235$-$1216}	  & 12:35:42.9$-$12:16:23 &	0.42 &	18.21	&	0.69	&	1162.1$\pm$1.7	&	33.3$\pm$3.9	&	0.33$\pm$0.04	&	 9.7	&	6.86	&	6.97	\\
\textbf{Dw\,1235$-$0215}	  & 12:35:53.0$-$02:15:54 &	0.82 &	18.80	&	0.46	&	1160.5$\pm$1.3	&	17.3$\pm$3.0	&	0.17$\pm$0.04	&	 9.6	&	6.56	&	6.38	\\
Dw\,1235+7050	              & 12:35:59.5+70:50:53	  &	0.57 &	18.59	&	0.67	&	1844.5$\pm$0.9	&	28.6$\pm$2.2	&	0.56$\pm$0.04	&	30.9	&	8.10	&	7.78	\\
dw\,1238$-$1122	          & 12:38:33.6$-$11:22:05 &	0.66 &	18.54	&	0.73	&	2319.0$\pm$1.0	&	24.6$\pm$2.3	&	0.47$\pm$0.05	&	33.2	&	8.08	&	7.96	\\
\textbf{PGC\,4074723}	  & 12:40:29.9+47:22:04	  &	0.70 &	17.55	&	0.56	&	 524.4$\pm$0.6	&	22.4$\pm$1.3	&	0.69$\pm$0.04	&	 8.4	&	7.05	&	6.91	\\
dw\,1245$-$1041	          & 12:45:07.2$-$10:41:56 &	0.74 &	18.84	&	0.65	&	1479.4$\pm$1.5	&	14.6$\pm$3.6	&	0.08$\pm$0.03	&	22.3	&	6.99	&	7.38	\\
Dw\,1247$-$0824            & 12:47:25.0$-$08:24:29 &	0.66 &	15.73	&	0.59	&	1214.1$\pm$1.3	&	86.4$\pm$2.7	&	3.04$\pm$0.10	&	14.8	&	8.19	&	8.18	\\
\textbf{PGC\,974136}	  & 12:49:59.5$-$10:45:22 &	0.40 &	16.23	&	0.65	&	 993.8$\pm$0.9	&	32.5$\pm$2.2	&	1.04$\pm$0.05	&	 7.9	&	7.19	&	7.52	\\
MKT\,J125$-$12	          & 12:52:25.4$-$12:43:05 &	0.62 &	16.89	&	0.36	&	1299.8$\pm$0.7	&	55.5$\pm$1.6	&	3.06$\pm$0.07	&	16.8	&	8.31	&	7.49	\\
\textbf{SMDG\,1256$-$00}  & 12:56:35.8$-$00:31:44 &	0.67 &	17.88	&	0.51	&	1177.2$\pm$0.8	&	32.5$\pm$2.0	&	0.72$\pm$0.04	&	10.2	&	7.24	&	6.87	\\
PGC\,044302	              & 12:57:27.7$-$12:06:28 &	0.63 &	15.88	&	0.55	&	1407.7$\pm$0.7	&	44.2$\pm$1.8	&	1.73$\pm$0.06	&	18.4	&	8.14	&	8.25	\\
\textbf{dw\,1303+42}        & 13:03:14.0+42:22:17	  &	0.63 &	18.28	&	0.60	&	 450.2$\pm$2.3	&	27.0$\pm$5.3	&	0.22$\pm$0.04	&	 6.4	&	6.33	&	6.44	\\
dw\,1305+41	              & 13:05:29.0+41:53:24	  &	0.88 &	17.02	&	0.41	&	1154.3$\pm$1.3	&	62.5$\pm$3.0	&	1.55$\pm$0.06	&	21.9	&	8.24	&	7.74	\\
KDG\,218	                  & 13:05:44.0$-$07:45:20 &	0.71 &	16.40	&	0.65	&	1474.9$\pm$1.6	&	22.3$\pm$3.7	&	0.28$\pm$0.03	&	22.7	&	7.52	&	8.37	\\
\textbf{dw\,1305+38}	      & 13:05:58.0+38:05:43	  &	0.70 &	18.32	&	0.56	&	 419.6$\pm$1.6	&	28.8$\pm$3.6	&	0.14$\pm$0.03	&	 5.3	&	5.96	&	6.21	\\
\textbf{Dw\,1311+4051}	  & 13:11:41.3+40:51:47	  &	0.61 &	18.44	&	0.43	&	 603.0$\pm$0.6	&	24.3$\pm$1.5	&	0.50$\pm$0.05	&	 9.3	&	7.01	&	6.45	\\
\textbf{dw\,1313+46}	      & 13:13:02.0+46:36:08	  &	0.83 &	17.96	&	0.34	&	 388.6$\pm$0.6	&	26.6$\pm$1.3	&	0.86$\pm$0.04	&	 6.1	&	6.87	&	6.14	\\
\textbf{CGCG\,189-050}	  & 13:17:04.9+37:57:08	  &	0.77 &	15.48	&	0.58	&	 333.6$\pm$0.5	&	30.1$\pm$1.1	&	2.05$\pm$0.06	&	 4.2	&	6.93	&	7.16	\\
\textbf{PGC\,2229803}	  & 13:27:53.1+43:48:55	  &	0.65 &	16.28	&	0.66	&	 436.2$\pm$0.7	&	30.5$\pm$1.6	&	0.62$\pm$0.05	&	 6.5	&	6.78	&	7.33	\\
\textbf{LV\,J1328+4937}	  & 13:28:31.2+49:37:37	  &	0.55 &	17.44	&	0.37	&	 402.3$\pm$0.5	&	26.8$\pm$1.1	&	0.99$\pm$0.04	&	 6.6	&	7.01	&	6.47	\\
\textbf{UGC\,A361}	      & 13:32:36.2+49:49:49	  &	0.59 &	16.76	&	0.71	&	 216.1$\pm$1.6	&	22.7$\pm$3.7	&	0.11$\pm$0.03	&	 3.7	&	5.57	&	6.73	\\
\textbf{MCG+08-25-28}	  & 13:36:44.8+44:35:57	  &	0.52 &	15.99	&	0.58	&	 485.7$\pm$0.7	&	32.6$\pm$1.6	&	1.00$\pm$0.05	&	 7.6	&	7.13	&	7.47	\\
Dw\,1337+6905	              & 13:37:43.2+69:05:42	  &	0.78 &	18.09	&	0.52	&	1705.0$\pm$0.8	&	39.4$\pm$1.9	&	1.00$\pm$0.04	&	29.2	&	8.30	&	7.72	\\
\textbf{Dw\,1339+39}	      & 13:39:45.1+39:08:09	  &	0.66 &	16.99	&	0.78	&	 674.1$\pm$0.5	&	24.0$\pm$1.1	&	0.80$\pm$0.04	&	 9.9	&	7.26	&	7.60	\\
dw\,1340+45	              & 13:40:37.0+45:41:54	  &	0.42 &	18.44	&	0.37	&	1385.8$\pm$1.0	&	30.7$\pm$2.3	&	0.70$\pm$0.04	&	25.0	&	8.01	&	7.23	\\
\textbf{Dw\,1341+4204}	  & 13:41:58.8+42:04:05	  &	0.77 &	19.45	&	0.45	&	 248.6$\pm$0.9	&	17.1$\pm$2.0	&	0.16$\pm$0.02	&	 3.7	&	5.70	&	5.27	\\
\textbf{LV\,J1342+4840}	  & 13:42:20.1+48:40:57	  &	0.69 &	16.52	&	0.58	&	 438.3$\pm$0.8	&	25.2$\pm$1.8	&	0.52$\pm$0.04	&	 7.4	&	6.82	&	7.23	\\
Dw\,1352+6142	              & 13:52:39.4+61:42:50	  &	0.56 &	17.44	&	0.58	&	2111.4$\pm$1.6	&	67.1$\pm$4.0	&	1.35$\pm$0.06	&	34.2	&	8.57	&	8.20	\\
Dw\,1355+0859	              & 13:55:58.7+08:59:39	  &	0.48 &	17.78	&	0.81	&	1219.6$\pm$3.0	&	44.7$\pm$7.0	&	0.38$\pm$0.05	&	18.2	&	7.47	&	7.85	\\
Dw\,1403+4924	              & 14:03:19.0+49:24:54	  &	0.72 &	18.94	&	0.54	&	2033.0$\pm$1.3	&	23.5$\pm$3.1	&	0.29$\pm$0.04	&	32.2	&	7.85	&	7.50	\\
Dw\,1418+4607	              & 14:18:31.5+46:07:51	  &	0.63 &	18.21	&	0.60	&	1825.3$\pm$1.4	&	26.5$\pm$3.3	&	0.42$\pm$0.04	&	28.6	&	7.91	&	7.76	\\
dw\,1446+58	              & 14:46:01.0+58:34:05	  &	0.65 &	18.80	&	0.59	&	2300.3$\pm$2.8	&	47.1$\pm$6.6	&	0.51$\pm$0.05	&	36.8	&	8.21	&	7.73	\\
Dw\,1459+44	              & 14:59:38.4+44:40:23	  &	0.95 &	16.09	&	0.47	&	 732.4$\pm$1.1	&	80.1$\pm$2.3	&	5.16$\pm$0.15	&	14.5	&	8.40	&	7.84	\\
\textbf{Dw\,1533+67}	      & 15:33:28.1+67:45:29	  &	0.32 &	17.57	&	0.54	&	 397.6$\pm$0.6	&	21.4$\pm$1.4	&	0.41$\pm$0.04	&	10.1	&	6.99	&	7.04	\\
Dw\,1539+4515	              & 15:39:25.9+45:15:04	  &	0.34 &	17.97	&	0.61	&	2605.6$\pm$2.5	&	60.6$\pm$5.9	&	0.66$\pm$0.05	&	40.3	&	8.40	&	8.17	\\
\textbf{Dw\,1559+46}	      & 15:59:02.6+46:23:40	  &	0.88 &	17.10	&	0.43	&	  76.7$\pm$0.4	&	28.5$\pm$1.0	&	1.42$\pm$0.04	&	 3.3	&	6.56	&	6.09	\\
Dw\,1608+4058	              & 16:08:19.4+40:58:12	  &	0.87 &	18.89	&	0.39	&	1982.1$\pm$1.0	&	28.5$\pm$2.5	&	0.48$\pm$0.04	&	30.9	&	8.03	&	7.25	\\
Dw\,1615+5422	              & 16:15:42.1+54:22:03	  &	0.82 &	18.09	&	0.38	&	3697.3$\pm$0.7	&	52.2$\pm$1.6	&	1.81$\pm$0.06	&	57.1	&	9.14	&	8.10	\\
Dw\,1633+6906	              & 16:33:01.4+69:06:04	  &	0.85 &	18.00	&	0.59	&	1122.9$\pm$1.1	&	35.8$\pm$2.5	&	0.85$\pm$0.05	&	19.6	&	7.88	&	7.50	\\
Dw\,1643+0749	              & 16:43:25.1+07:49:30   &	0.30 &	18.84	&	0.51	&	1434.6$\pm$1.5	&	18.4$\pm$3.5	&	0.19$\pm$0.03	&	19.6	&	7.23	&	7.07	\\
VV\,507	                  & 17:00:42.2+53:21:36	  &	0.81 &	16.00	&	0.41	&	1122.6$\pm$0.3	&	46.0$\pm$0.7	&	3.39$\pm$0.05	&	20.0	&	8.50	&	8.07	\\
Dw\,1709+74	              & 17:09:45.6+74:10:44	  &	0.81 &	17.17	&	0.97	&	1325.4$\pm$2.9	&	48.3$\pm$7.0	&	0.41$\pm$0.05	&	21.8	&	7.65	&	8.48	\\
Dw\,1721+7140	              & 17:21:45.4+71:40:59	  &	0.68 &	17.25	&	0.60	&	1076.1$\pm$0.6	&	36.2$\pm$1.4	&	1.58$\pm$0.05	&	18.6	&	8.11	&	7.77	\\
\textbf{Dw\,1735+57}	      & 17:35:34.6+57:48:47   &	0.61 &	16.75	&	0.53	&	  42.9$\pm$0.4	&	25.7$\pm$1.0	&	0.90$\pm$0.03	&	 4.4	&	6.61	&	6.64	\\

\end{longtable*}
\end{ThreePartTable}

\addtocounter{figure}{-1}
\begin{figure*}[ht]
\centering
\caption{Continue}
\includegraphics[width=1\linewidth]{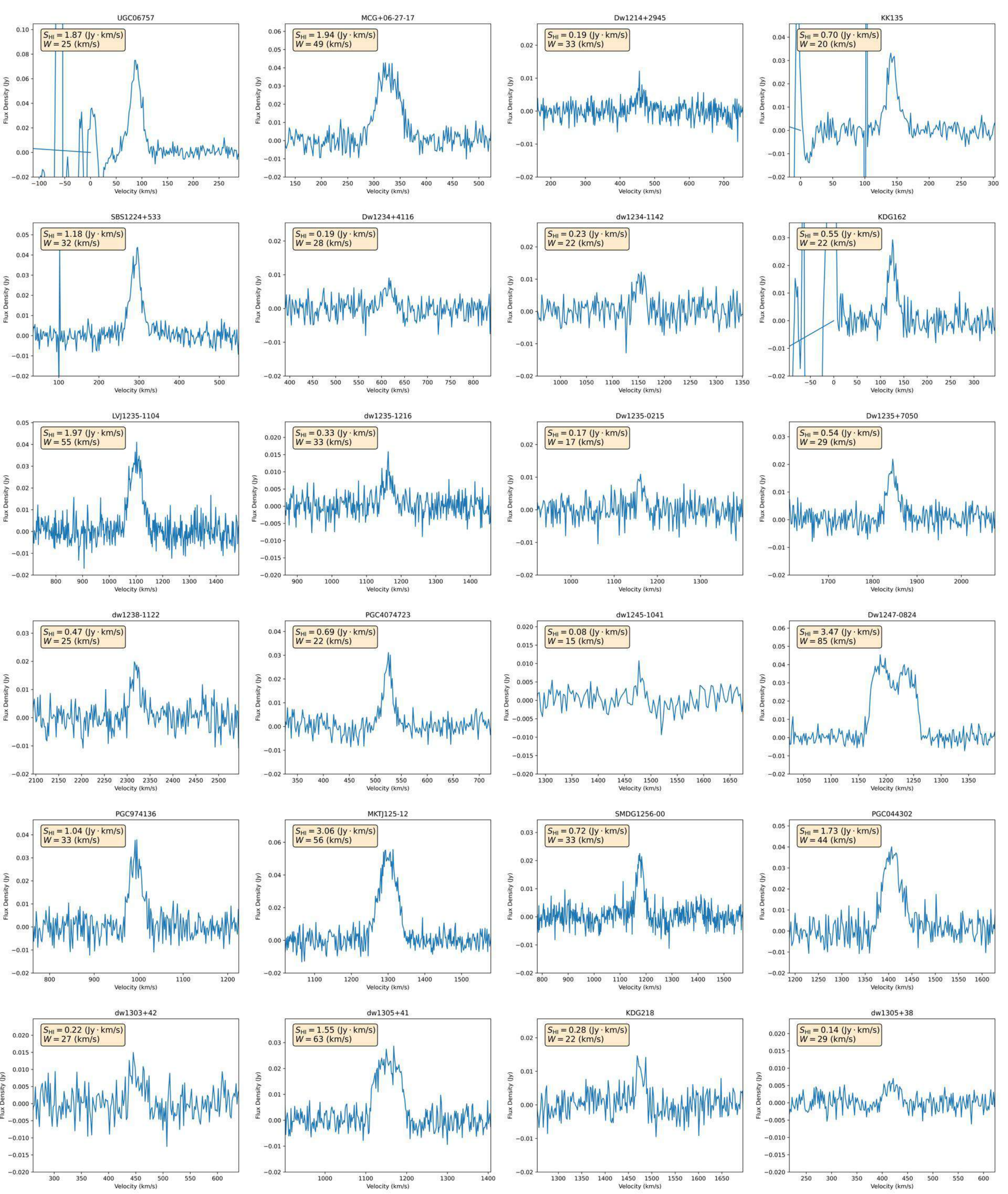}
\end{figure*}
\addtocounter{figure}{-1}
\begin{figure*}[ht]
\centering
\caption{Continue}
\includegraphics[width=1\linewidth]{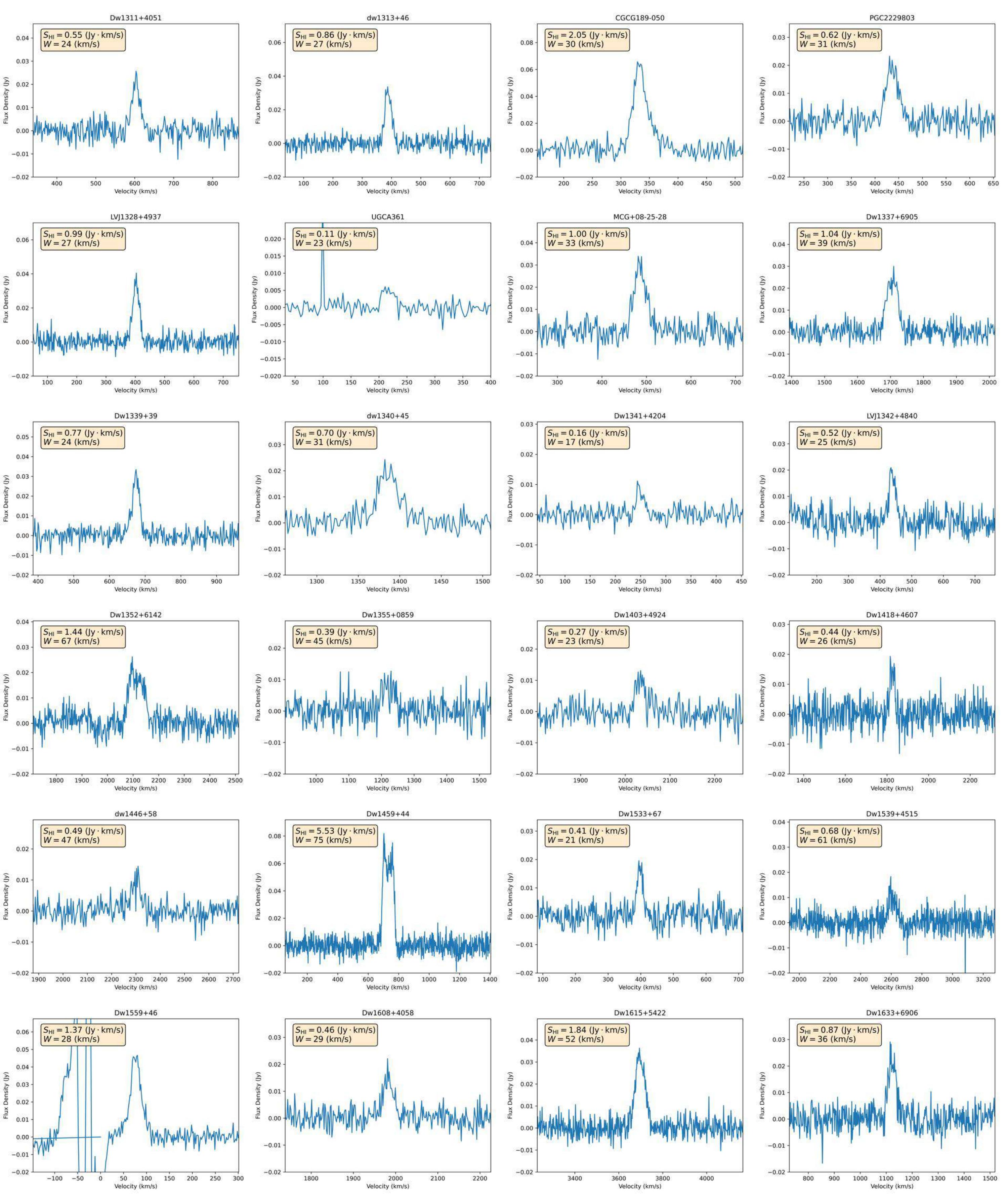}
\end{figure*}
\addtocounter{figure}{-1}
\begin{figure*}[ht]
\centering
\caption{Continue}
\includegraphics[width=1\linewidth]{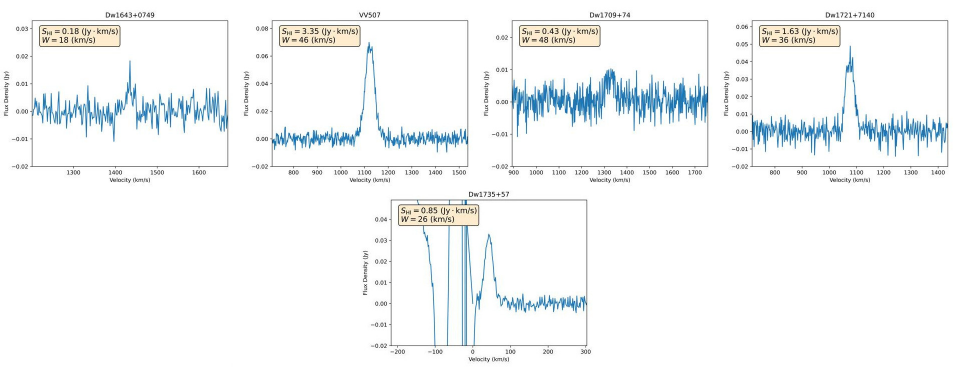}
\end{figure*}

\begin{figure*}
\centering
\caption{Images of 28 undetected nearby late-type dwarf galaxies taken from DESI Legacy Imaging Surveys (\textit{KKR56} from PanSTARRS). Each image side is $2\arcmin$, North is to the top and East is to the left.}
\includegraphics[width=\textwidth]{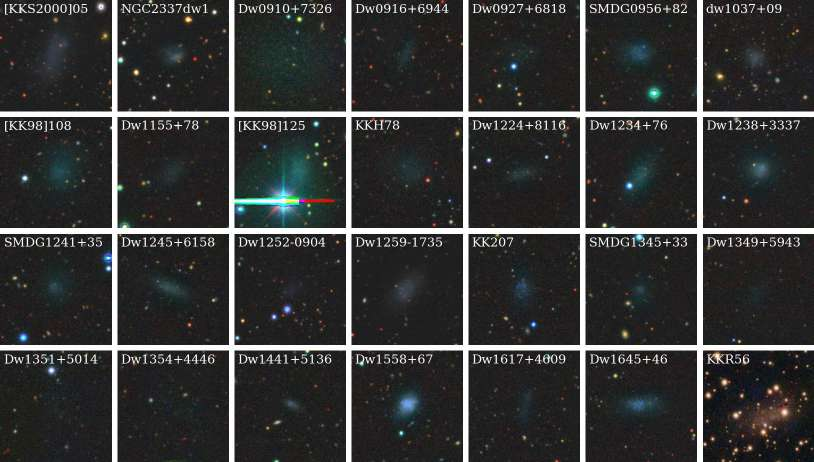}
\label{fig:mosaic28}
\end{figure*}

The remaining 28 undetected galaxies are presented in Table~\ref{table2}, where the column designations are the same as for the first five columns of Table~\ref{table1}. Their Legacy Survey images are shown in Fig.~\ref{fig:mosaic28} (except KKR56, which is from from Pan-STARRS1). The size $2\arcmin\times2\arcmin$ and orientation (North up and East to the left) are the same as in Fig.~\ref{fig:mosaic77}.

It should be noted that the majority of our targets belong to the late morphological types:
irregular (Irr), Magellanic (Im), blue compact dwarf (BCD), rich in neutral hydrogen. 
Some dwarf galaxies are classified as Tr-type, a transitional type between irregular and spheroidal (dSph) dwarfs. 
Along with dwarf galaxies of the late types, we included in the target list only one nearby dSph object, Dw\,0910+7326, in order to check the presence of neutral hydrogen in it. 
According to subsequent optical spectral observations~\citep{2025MNRAS.537L..21K}, five of the undetected galaxies, Dw\,1234+76, Dw\,1245+6158, Dw\,1558+67, Dw\,1645+46 and SMDG\,0956+82, are low-velocity objects, whose HI line is confused with local Galactic hydrogen.

Table~\ref{table3} contains a summary of our GBT targets that have independent radial velocity measurements obtained with the 500-m FAST radio telescope~\citep{zha2024, 2024A&A...684L..24K} as well as with optical measurements collected in the current version of the UNGC catalog~\citep{kar2013}. Our measurements are in excellent agreement with the FASHI survey in terms of velocity: $\langle V^\mathrm{FAST} - V^\mathrm{GBT}\rangle = -0.4\pm0.7$~km\,s$^{-1}$ with standard deviation of $\sigma_{V} = 3.7$~km\,s$^{-1}$; HI line width at the 50\% level: $\langle W_\mathrm{50}^\mathrm{FAST} - W_\mathrm{50}^\mathrm{GBT}\rangle = 0.9\pm0.8$~km\,s$^{-1}$ and $\sigma_{W_{50}}=4.3$~km\,s$^{-1}$; and total flux: $\langle S_\mathrm{HI}^\mathrm{FAST} - S_\mathrm{HI}^\mathrm{GBT}\rangle = -0.09\pm0.06$~Jy\,km\,s$^{-1}$ and $\sigma_S=0.33$~Jy\,km\,s$^{-1}$. Comparison of the GBT measurements with data collected in the LV database gives $\langle V^\mathrm{UNGC} - V^\mathrm{GBT}\rangle = -19\pm10$~km\,s$^{-1}$ with $\sigma = 59$~km\,s$^{-1}$, which turns out to be noticeably worse due to the low accuracy of optical velocity measurements.

\setlength{\tabcolsep}{3.5pt}
\begin{table}
\centering
\caption{List of galaxies not detected with GBT.}
\label{table2}
\begin{tabular}{l c c c c} 

\hline\hline
\multicolumn{1}{c}{Name}   &     RA(2000.0) Dec     &  $b/a$  &           $B$             &         $B - V$           \\

                             &                        &         & \multicolumn{1}{c}{mag}  & \multicolumn{1}{c}{mag}  \\

\hline

[KKS2000]\,05                &	02:49:26.1$-$13:12:42 &	 0.43	&	17.62	&	0.59	\\
NGC\,2337dw1	             &	07:08:29.7+44:37:26	  &	 0.77	&	18.75	&	0.77	\\
Dw\,0910+7326	             &	09:10:15.6+73:26:24	  &	 0.88	&	17.40	&	1.29	\\
Dw\,0916+6944	             &	09:16:43.7+69:44:01	  &	 0.33	&	19.75	&	0.71	\\
Dw\,0927+6818              	 &	09:27:27.8+68:18:55	  &	 0.76	&	19.43	&	0.88	\\
SMDG\,0956+82	             &	09:56:13.0+82:53:24	  &	 0.60	&	17.99	&	0.67	\\
Dw\,1037+09	                 &	10:37:40.6+09:06:20	  &	 0.77	&	18.90	&	0.82	\\
\lbrack KK98\rbrack\,108	 &	11:40:03.6+46:28:43	  &	 0.84	&	17.70	&	0.77	\\
Dw\,1155+78	                 &	11:55:54.2+78:04:44	  &	 0.73	&	19.17	&	0.65	\\
\lbrack KK98\rbrack\,125	 &	12:12:41.9+68:55:38	  &	 0.87	&   17.40   &	  	    \\
KKH78	                     &	12:17:44.5+33:20:43	  &	 0.63	&	18.29	&	0.73	\\
Dw\,1224+8116	             &	12:24:49.7+81:16:59	  &	 0.44	&	19.36	&	0.83	\\
Dw\,1234+76	                 &	12:34:23.3+76:43:34	  &	 0.58	&	17.97	&	0.75	\\
Dw\,1238+3337	             &	12:38:18.0+33:37:59	  &	 0.92	&	17.66	&	0.67	\\
SMDG\,1241+35	             &	12:41:11.0+35:11:46	  &	 0.92	&	18.73	&	0.75	\\
Dw\,1245+6158	             &	12:45:49.0+61:58:08	  &	 0.49	&	18.42	&	0.72	\\
Dw\,1252$-$0904	             &	12:52:03.4$-$09:04:26 &	 0.70	&	19.62	&	0.35	\\
Dw\,1259$-$1735	             &	12:59:40.3$-$17:35:46 &	 0.53	&	18.11	&	0.63	\\
KK\,207   	                 &	13:33:25.7+56:30:00	  &	 0.70	&	18.38	&	0.48	\\
SMDG\,1345+33 	             &	13:45:11.0+33:11:31	  &	 0.86	&	19.23	&	0.73	\\
Dw\,1349+5943	             &	13:49:50.9+59:43:44	  &	 0.90	&	20.13	&	0.85	\\
Dw\,1351+5014	             &	13:51:59.0+50:14:53	  &	 0.82	&	20.50   &	  	    \\
Dw\,1354+4446	             &	13:54:10.1+44:46:48	  &	 0.80	&	  	    &	  	    \\
Dw\,1441+5136	             &	14:41:56.6+51:36:40	  &	 0.57	&	19.56	&	0.69	\\
Dw\,1558+67	                 &	15:58:46.8+67:51:29	  &	 0.87	&	16.75	&	0.47	\\
Dw\,1617+4609	             &	16:17:35.8+46:09:32	  &	 0.48	&	19.28	&	0.47	\\
Dw\,1645+46	                 &	16:45:48.5+46:47:24	  &	 0.67	&	17.61	&	0.52	\\
KKR56	                     &	20:48:24.1+58:37:06	  &	 0.64	&	17.60   &	 	    \\
\hline\hline
\end{tabular}
\raggedright\footnotesize{The upper limit of the flux for the undetected sources is about~0.09~Jy~km~s$^{-1}$, which corresponds to $m_{21} = 20\fm0$.}
\end{table}

\setlength{\tabcolsep}{4.8pt}
\begin{table*}

\caption{Comparison of radial velocity and HI measurements.} \label{table3} 
\centering
\begin{tabular}{l c r c c r c c r} 

\hline\hline
                                     & 
                                     &
\multicolumn{3}{c}{GBT}             &               
\multicolumn{3}{c}{FASHI}           &      
\multicolumn{1}{c}{UNGC}            \\

\cline{3-9}

\multicolumn{1}{c}{Name}            & 
\multicolumn{1}{c}{RA(2000.0) Dec}   &
\multicolumn{1}{c}{$V_\mathrm{h}$}   & 
\multicolumn{1}{c}{$S_\mathrm{HI}$}  &
\multicolumn{1}{c}{$W_\mathrm{50}$}  &   
\multicolumn{1}{c}{$V_\mathrm{h}$}   &  
\multicolumn{1}{c}{$S_\mathrm{HI}$}  &
\multicolumn{1}{c}{$W_\mathrm{50}$}  & 
\multicolumn{1}{c}{$V_\mathrm{h}$}   \\

                                      & 
                                      &
\multicolumn{1}{c}{km s$^{-1}$}      & 
\multicolumn{1}{c}{Jy km s$^{-1}$}   &
\multicolumn{1}{c}{km s$^{-1}$}      &   
\multicolumn{1}{c}{km s$^{-1}$}      &  
\multicolumn{1}{c}{Jy km s$^{-1}$}   &
\multicolumn{1}{c}{km s$^{-1}$}      & 
\multicolumn{1}{c}{km s$^{-1}$}      \\

\hline
PGC\,025409	    &	09:02:50.6+71:18:22	 &	 408.4$\pm$0.5	&\ 		 	        &\                  &\ 			        &\ 			        &\ 			        &	 416$\pm$10	\\
UGC\,4918	    &	09:19:17.7+69:48:04	 &	 723.3$\pm$1.5	&\ 			        &\ 			        &\ 			        &\ 			        &\ 			        &	 710$\pm$41	\\
KJ78	        &	09:20:36.4+49:40:31	 &	 603.8$\pm$0.4	&	0.89$\pm$0.04	&	22.0$\pm$0.9	&	606.1$\pm$0.6	&	0.74$\pm$0.06	&	22.4$\pm$1.2	&	 477$\pm$30	\\
MCG+09-16-10    &	09:23:17.0+51:58:22	 &	 485.5$\pm$0.7	&	3.09$\pm$0.06	&	57.5$\pm$1.5	&	485.5$\pm$0.5	&	3.80$\pm$0.13	&	65.8$\pm$1.0	&	 484$\pm$4  \\
LV\,J0935$-$1348&	09:35:21.6$-$13:48:52&	 811.3$\pm$1.7	&\ 			        &\ 	   		        &\ 			        &\ 			        &\ 			        &	 796$\pm$45	\\
PGC\,154449	    &	09:57:08.9$-$09:15:48&	 562.7$\pm$1.8	&\ 			        &\ 	   		        &\ 			        &\ 			        &\ 			        &	 543$\pm$45	\\
LV\,J1000+5022  &	10:00:25.5+50:22:45	 &	 548.7$\pm$1.2	&	0.84$\pm$0.05	&	48.7$\pm$2.8	&	551.1$\pm$1.2	&	0.73$\pm$0.08	&	52.6$\pm$2.4	&	 538$\pm$11	\\
Dw\,1012+4259	&	10:12:42.7+42:59:31	 &	2297.0$\pm$1.7	&	0.74$\pm$0.04	&	49.5$\pm$4.2	&	2296.0$\pm$1.8	&	0.53$\pm$0.08	&	52.9$\pm$3.5	&	2326$\pm$47	\\
KUG\,1013+414	&	10:16:15.6+41:09:58	 &	 506.6$\pm$3.2	&\ 			        &\ 	   		        &\ 			        &\ 			        &\ 			        &	 518$\pm$1	\\
PGC\,30114	    &	10:18:43.0+46:02:44	 &	 586.6$\pm$0.6	&	5.12$\pm$0.07	&	55.2$\pm$1.3	&	584.5$\pm$0.2	&	5.61$\pm$0.10	&	59.5$\pm$0.4	&	 566$\pm$256\\
LV\,J1028+4240  &	10:28:33.0+42:40:07	 &	 559.0$\pm$0.6	&	1.83$\pm$0.07	&	34.2$\pm$1.3	&	557.2$\pm$0.4	&	1.70$\pm$0.07	&	30.2$\pm$0.8	&	 547$\pm$5	\\
PGC\,2277751    &	10:35:12.1+46:14:12	 &	 544.0$\pm$1.9	&\ 			        &\ 	   		        &\ 			        &\ 			        &\ 			        &	 505$\pm$23	\\
KUG\,1033+366B  &	10:36:17.6+36:25:31	 &	 618.2$\pm$0.9	&	0.70$\pm$0.04	&	34.7$\pm$2.1	&	619.1$\pm$0.7	&	0.75$\pm$0.07	&	32.2$\pm$1.5	&	 620$\pm$2	\\
VV747	        &	10:57:47.0+36:15:39	 &	 630.2$\pm$0.7	&	4.91$\pm$0.08	&	75.4$\pm$1.6	&	628.3$\pm$0.4	&	5.92$\pm$0.15	&	84.8$\pm$0.8	&	 619$\pm$2	\\
SMDG\,1103+60   &	11:03:56.4+60:29:53	 &	1060.7$\pm$1.6	&	0.38$\pm$0.05	&	27.6$\pm$3.7	&	1059.9$\pm$1.2	&	0.33$\pm$0.06	&	30.4$\pm$2.3	&			    \\
UGC\,6451	    &	11:28:46.4+79:36:07	 &	  49.6$\pm$0.3	&\ 			        &\ 	   		        &\ 			        &\ 			        &\ 			        &	  33$\pm$20	\\
UGC\,06757	    &	11:46:59.1+61:20:05	 &	  88.4$\pm$0.3	&\ 			        &\ 	   		        &\ 			        &\ 			        &\ 			        &	  82$\pm$4	\\
MCG+06-27-17    &	12:09:56.4+36:26:07	 &	 326.8$\pm$0.8	&	1.94$\pm$0.07	&	49.3$\pm$1.9	&	325.5$\pm$1.2	&	1.56$\pm$0.14	&	47.6$\pm$2.4	&			    \\
Dw\,1214+2945	&	12:14:26.6+29:45:50	 &	 457.5$\pm$2.7	&\ 			        &\ 	   		        &\ 			        &\ 			        &\ 			        &	 426$\pm$69	\\
KK135	        &	12:19:34.7+58:02:34	 &	 142.5$\pm$0.5	&\ 			        &\ 	   		        &\ 			        &\ 			        &\ 			        &	 215$\pm$40	\\
SBS1224+533	    &	12:26:52.6+53:06:19	 &	 292.1$\pm$0.5	&	1.18$\pm$0.04	&	31.6$\pm$1.2	&	291.4$\pm$0.5	&	1.50$\pm$0.08	&	31.0$\pm$0.9	&	 300$\pm$3	\\
Dw\,1234+4116	&	12:34:38.2+41:16:34	 &	 614.3$\pm$1.4	&\ 			        &\ 	   		        &\ 			        &\ 			        &\ 			        &	 601$\pm$3	\\
KDG\,162	    &	12:35:01.6+58:23:08	 &	 125.3$\pm$0.8	&\ 			        &\ 	   		        &\ 			        &\ 			        &\ 			        &	 105$\pm$50	\\
LV\,J1235$-$1104&	12:35:39.4$-$11:04:01&	1099.7$\pm$0.9	&\ 			        &\ 	   		        &\ 			        &\ 			        &\ 			        &	1110$\pm$10	\\
PGC\,4074723    &	12:40:29.9+47:22:04	 &	 524.4$\pm$0.6	&	0.69$\pm$0.04	&	22.4$\pm$1.3	&	522.8$\pm$0.6	&	0.55$\pm$0.04	&	20.1$\pm$1.2	&	 229$\pm$132\\
Dw\,1247$-$0824 &	12:47:25.0$-$08:24:29&	 524.4$\pm$0.6	&\ 			        &\ 	   		        &\ 			        &\ 			        &\ 			        &	1215$\pm$10	\\
dw\,1303+42	    &	13:03:14.0+42:22:17	 &	 450.2$\pm$2.3	&	0.22$\pm$0.04	&	27.0$\pm$5.3	&	448.9$\pm$0.9	&	0.33$\pm$0.03	&	32.7$\pm$1.9	&			    \\
Dw\,1311+4051	&	13:11:41.3+40:51:47	 &	 603.0$\pm$0.6	&	0.50$\pm$0.05	&	24.3$\pm$1.5	&	604.2$\pm$0.8	&	0.52$\pm$0.07	&	21.3$\pm$1.7	&			    \\
dw\,1313+46	    &	13:13:02.0+46:36:08	 &	 388.6$\pm$0.6	&	0.86$\pm$0.04	&	26.6$\pm$1.3	&	388.5$\pm$0.5	&	0.95$\pm$0.06	&	26.9$\pm$1.1	&			    \\
CGCG\,189-050	& 13:17:04.9+37:57:08	 &	 333.6$\pm$0.5	&	2.05$\pm$0.06	&	30.1$\pm$1.1	&	333.5$\pm$0.4	&	2.27$\pm$0.07	&	27.1$\pm$0.7	&	 324$\pm$6	\\
PGC\,2229803	&	13:27:53.1+43:48:55	 &	 436.2$\pm$0.7	&	0.62$\pm$0.05	&	30.5$\pm$1.6	&	433.9$\pm$0.8	&	0.64$\pm$0.07	&	29.9$\pm$1.7	&	 452$\pm$39	\\
LV\,J1328+4937  &	13:28:31.2+49:37:37	 &	 402.3$\pm$0.5	&	0.99$\pm$0.04	&	26.8$\pm$1.1	&	400.8$\pm$0.6	&	0.96$\pm$0.07	&	21.7$\pm$1.2	&	 395$\pm$9	\\
MCG+08-25-28    &	13:36:44.8+44:35:57	 &	 485.7$\pm$0.7	&	1.00$\pm$0.05	&	32.6$\pm$1.6	&	487.9$\pm$0.8	&	1.04$\pm$0.08	&	30.3$\pm$1.5	&			    \\
Dw\,1339+39	    &	13:39:45.1+39:08:09	 &	 674.1$\pm$0.5	&	0.80$\pm$0.04	&	24.0$\pm$1.1	&	681.9$\pm$0.7	&	1.03$\pm$0.07	&	25.4$\pm$1.4	&			    \\
dw\,1340+45	    &	13:40:37.0+45:41:54	 &	1385.8$\pm$1.0	&	0.70$\pm$0.04	&	30.7$\pm$2.3	&	1375.9$\pm$1.8	&	0.79$\pm$0.11	&	36.3$\pm$3.5	&			    \\
LV\,J1342+4840  &	13:42:20.1+48:40:57	 &	 438.3$\pm$0.8	&	0.52$\pm$0.04	&	25.2$\pm$1.8	&	434.5$\pm$0.8	&	0.60$\pm$0.05	&	22.5$\pm$1.6	&	 438$\pm$2	\\
Dw\,1352+6142	&	13:52:39.4+61:42:50	 &	2111.4$\pm$1.6	&	1.35$\pm$0.06	&	67.1$\pm$4.0	&	2111.2$\pm$0.9	&	1.75$\pm$0.11	&	66.2$\pm$1.7	&			    \\
Dw\,1403+4924	&	14:03:19.0+49:24:54	 &	2033.0$\pm$1.3	&	0.29$\pm$0.04	&	23.5$\pm$3.1	&	2022.5$\pm$1.7	&	0.38$\pm$0.06	&	28.0$\pm$3.3	&			    \\
Dw\,1418+4607	&	14:18:31.5+46:07:51	 &	1825.3$\pm$1.4	&	0.42$\pm$0.04	&	26.5$\pm$3.3	&	1829.5$\pm$1.1	&	0.43$\pm$0.07	&	32.1$\pm$2.1	&			    \\
dw\,1446+58	    &	14:46:01.0+58:34:05	 &	2300.3$\pm$2.8	&	0.51$\pm$0.05	&	47.1$\pm$6.6	&	2299.4$\pm$1.1	&	0.41$\pm$0.05	&	40.9$\pm$2.3	&			    \\
Dw\,1459+44	    &	14:59:38.4+44:40:23	 &	 732.4$\pm$1.1	&	5.16$\pm$0.15	&	80.1$\pm$2.3	&	732.8$\pm$0.2	&	5.86$\pm$0.11	&	80.8$\pm$0.5	&	 683$\pm$15	\\
Dw\,1539+4515	&	15:39:25.9+45:15:04	 &	2605.6$\pm$2.5	&	0.66$\pm$0.05	&	60.6$\pm$5.9	&	2607.5$\pm$1.4	&	0.53$\pm$0.07	&	69.5$\pm$2.9	&			    \\
Dw\,1559+46	    &	15:59:02.6+46:23:40	 &	  76.7$\pm$0.4	&\ 			        &\ 	   		        &\ 			        &\ 			        &\ 			        &	  67$\pm$10	\\
Dw\,1615+5422	&	16:15:42.1+54:22:03	 &	3697.3$\pm$0.7	&	1.81$\pm$0.06	&	52.2$\pm$1.6	&	3703.8$\pm$0.3	&	1.18$\pm$0.03	&	47.3$\pm$0.6	&			    \\
Dw\,1709+74	    &	17:09:45.6+74:10:44	 &	1325.4$\pm$2.9	&\ 			        &\ 	   		        &\ 			        &\ 			        &\ 			        &	1298$\pm$10	\\
Dw\,1735+57	    &	17:35:34.6+57:48:47	 &	  42.9$\pm$0.4	&\ 			        &\ 			        &\ 			        &\ 			        &\ 			        &	  42$\pm$10	\\
\hline\hline
\end{tabular}
\end{table*}

\section{Notes on individual objects}
\label{sec:Notes}

{\it Dw\,0139+1433.} According to~\cite{Carlsten_2022}, its distance is $D = 10.82$ Mpc. This dwarf is a probable satellite of the spiral galaxy NGC\,628 or NGC\,660.

{\it Dw\,0142+1317.} A probable companion to the ring-like spiral galaxy NGC\,660.

{\it Dw\,0214+2836.} Sb galaxy NGC\,865 with a radial velocity of $V_h= 2995$~km\,s$^{-1}$ is located $27\arcmin$ South.

{\it SMDG\,0223$-$02.} A probable satellite of UGC\,1862.

{\it [KKS\,2000]05.} Isolated irregular dwarf with smooth structure and a knot on the NW side.

{\it SMDG\,0740+40.} It is probably associated with a brighter dwarf, DDO\,46, having the distance $D = 10.4$~Mpc via TRGB.

{\it PGC\,025409.} Isolated dIrr galaxy with granular structure at the center.

{\it Dw\,0910+7326.} A new spheroidal dwarf near the M\,81 group discovered by \citet{kar2022} with a distance of 3.21~Mpc~\citep{cas2023}. 

{\it Dw\,0916+6944.} Isolated dIrr with granular structure or a distant spiral galaxy of low surface brightness.  

{\it UGC\,4918.} A satellite of SB0 galaxy NGC\,2787 that has a radial velocity of $V_h= 700$~km\,s$^{-1}$.

{\it Dw\,0927+6818.} An object of low surface brightness, a probable satellite of NGC\,2787.

{\it Dw\,1012+4259.} A probable satellite of S0a galaxy IC\,598 that has $V_h = 2245$~km\,s$^{-1}$.
   
{\it Dw\,1109+5447.} A new dIrr satellite of Scd galaxy NGC\,3556 that has $V_h= 696$~km\,s$^{-1}$ and TF-distance of 9.9~Mpc (UNGC).

{\it UGC\,6451.} A new remote member of the M\,81 group, being in contact with a distant spiral galaxy UGC\,6450.

{\it Dw\,1214+2945.}  A dwarf galaxy of smooth structure located in the region of Coma-I group, which is characterized with a high negative peculiar velocity~\citep{aba2009}. 

{\it Dw\,1234+76.}  An isolated dwarf with granular structure and an optical velocity $V_h = 202\pm76 $~km\,s$^{-1}$~\citep{2025MNRAS.537L..21K}.
  
{\it Dw\,1234+4116.}  A blue compact dwarf , a satellite of the interacting galaxy pair NGC\,4490/85.

{\it dw\,1234$-$1142, LV\,J1235$-$1104, dw1235$-$1216, PGC\,974136.} These are likely members of the galaxy group around NGC\,4594 (Sombrero).

{\it KDG\,162.}  A blue semi-resolved dwarf, probably associated with Sm-galaxy NGC\,4605 that has $V_h= 151$~km\,s$^{-1}$  and TRGB-distance of 5.55~Mpc (UNGC).

{\it Dw\,1235$-$0215.} dIrr galaxy with an asymmetric shape and granular structure as seen in Hyper Suprime-Cam (HSC) archive of the Subary telescope. It is probably associated with NGC\,4594 (Sombrero) that has $V_h=1090$~km\,s$^{-1}$ and TRGB-distance of  $9.55$~Mpc.

{\it Dw\,1238+3337.} An isolated blue dwarf with smoothed structure. Its optical velocity is 
 $V_h= 1083\pm10$~km\,s$^{-1}$~\citep{2025MNRAS.537L..21K}.

{\it SMDG\,1241+35.} Its optical velocity is $V_h=661\pm10$~km\,s$^{-1}$~\citep{2025MNRAS.537L..21K}. A probable satellite of NGC\,4631.
 
{\it Dw\,1245+6158.} Its optical velocity is $V_h= 68\pm10$~km\,s$^{-1}$~\citep{2025MNRAS.537L..21K}. A probable satellite of NGC\,4605.
  
{\it Dw\,1247$-$0824.} A blue compact dwarf. Based on its double-horned HI-line, it is more likely a member of the Virgo Southern Extension than a companion to the Sombrero galaxy.

{\it SMDG\,1256$-$00.} A possible peripheral member of the Virgo cluster.

{\it KDG\,218.} An Ultra-Diffuse Dwarf galaxy at a distance of 12.59 Mpc via TRGB~\citep{Karachentsev_2018}. 

{\it Dw\,1305+38.} A probable peripheral companion to the spiral galaxy NGC 5055.
  
{\it Dw\,1311+4051.} A new satellite of the galaxy M\,63 = NGC\,5055 that has $V_h=500$~km\,s$^{-1}$ and TRGB-distance of 9.04~Mpc.

{\it UGCA\,361.} Based on its low radial velocity (216~km\,s$^{-1}$) and corresponding kinematic distance of 4.0 Mpc, this is an isolated dwarf located in front of the M\,51 group and classified as Tr type.

{\it KK\,207.} An irregular dwarf with granular structure.
  
{\it Dw\,1339+39.} An isolated irregular dwarf. At a separation of $46\arcmin$ to NE from it, there is another dIrr galaxy DDO\,182 with $V_h= 663$~km\,s$^{-1}$.

{\it  Dw\,1341+4204.} Possibly a new peripheral member of the M\,101/M\,63/M\,51 filament. It is of a very low surface brightness (see HSC archive) and HI-rich.

{\it Dw\,1351+5014.} \citet{kara2023} discovered it as `a dwarf galaxy [...] of very low surface brightness, which is barely visible in the SDSS survey'.
However, deep $g$-band images from the Hyper Suprime-Cam Legacy Archive~\citep{2021PASJ...73..735T} revealed the absence of a real dwarf galaxy there.

{\it Dw\,1352+6142.} A probable member of a group around the spiral galaxy UGC\,8822. 
  
{\it Dw\,1354+4446.} It was found in the images of the DESI Legacy Surveys as a very low surface brightness object, however more deep images of the Hyper Suprime-Cam Subaru Strategic Program DR2~\citep{2019PASJ...71..114A} show no real nearby dwarf galaxy there.

{\it Dw\,1403+4924.} It is $16\arcmin$ away from Sa galaxy NGC\,5448 that has $V_h= 2016$~km\,s$^{-1}$ .

{\it Dw\,1418+4607.} Isolated blue compact dwarf projected into a distant pair of galaxies.
  
{\it dw\,1446+58.}  A probable member of a scattered group of galaxies around Sb galaxy NGC\,5777 that has $V_h= 2139$~km\,s$^{-1}$ .

{\it  Dw\,1459+44.} A blue compact dwarf with a large faint halo (see the HSC archive). At a projected separation of $16\arcmin$ from it to the East, there is another BCD galaxy UGC\,9660 with $V_h= 608$~km\,s$^{-1}$ and TRGB-distance of 10.76~Mpc (UNGC).

{\it Dw\,1533+67.} An isolated dIrr. At a projected separation of $55\arcmin$ to SE from it, there is another dIrr galaxy UGC\,9992 with $V_h= 423$~km\,s$^{-1}$  and TRGB-distance of 11.27~Mpc (UNGC).

{\it  Dw\,1539+4515.} An isolated dIrr with patchy structure.
 
{\it Dw\,1558+67.} An isolated blue compact dwarf with granular structure and the optical velocity of $V_h= -25\pm10$~km\,s$^{-1}$~\citep{2025MNRAS.537L..21K}.  The lack of HI emission may be caused by confusion of the HI-line from the galaxy with the local Milky Way emission.

{\it  Dw\,1559+46.} An isolated blue compact dwarf with granular structure.
 
{\it  Dw\,1608+4058.} An isolated dIrr with smooth structure. Possibly associated with a group around Sab galaxy UGC\,10200 = Mrk\,1104 having $V_h= 1991$~km\,s$^{-1}$.

{\it Dw\,1615+5422.} A peculiar spiral or an interacting binary system. Bright in the $FUV$-band.
  
{\it Dw\,1617+4609.} An isolated dIrr galaxy of low surface brightness.
  
{\it  Dw\,1645+46.} An isolated dIrr galaxy semi-resolved into stars, with the optical velocity of $V_h= -21\pm10$~km\,s$^{-1}$~\citep{2025MNRAS.537L..21K}. Its HI-line is masked by the hydrogen emission from our Galaxy.

{\it  VV\,507= I\,Zw171.} Peculiar blue patchy galaxy. Surprisingly, this bright object still had no measured radial velocity. At $59\arcmin$ to the West from it, there is a dwarf Sm galaxy DDO\,206= UGC\,10608 with $V_h= 1090$~km\,s$^{-1}$.

{\it  Dw\,1735+57.}  An isolated blue compact dwarf with granular structure and a negative optical velocity.
 
\begin{figure}
    \centering
    \includegraphics[width=\linewidth]{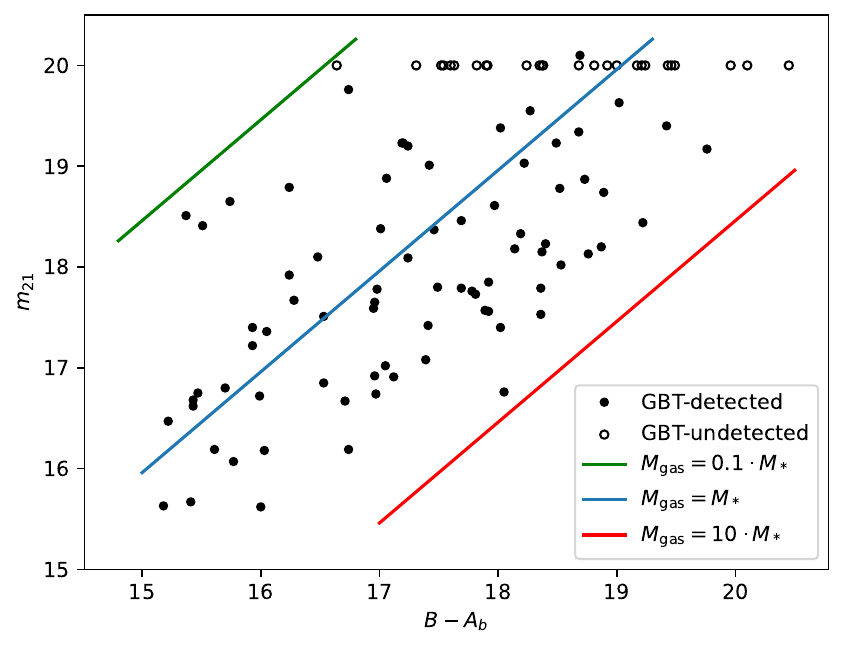}
    \includegraphics[width=\linewidth]{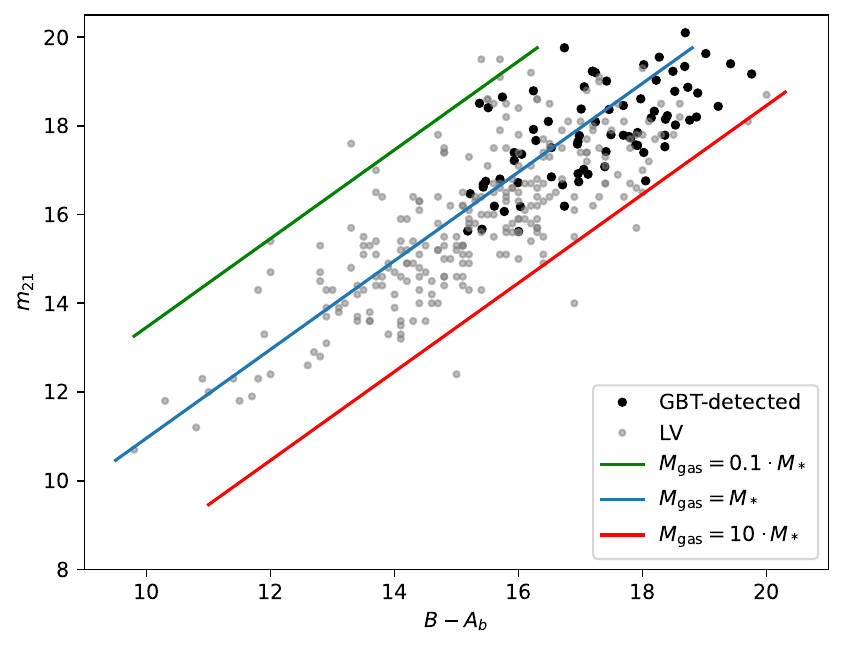}
    \caption{{\it Upper panel.}  Distribution of 105 observed dwarf galaxies according to $m_{21}$ magnitude and $B$-magnitude corrected for Galactic extinction. Three diagonal lines indicate  a ratio  $\mu = M_{\rm gas}/M_*$ equal to 0.1, 1, and 10. For undetected galaxies the $m_{21} = 20\fm0$ is assigned. {\it Bottom panel.}  Distribution of 77 GBT-detected dwarf galaxies according to $m_{21}$ magnitude and $B$-magnitude corrected for Galactic extinction combined with 266 late-type (T= 9, 10) dwarf galaxies in the Local Volume with TRGB distance estimates. Three diagonal lines correspond to $\mu$ equal to 0.1, 1, and 10.} \label{figure3}
\end{figure}

\section{Discussion}
\label{sec:Discussion}

The sample of 77 HI-detected galaxies is characterized by the following median values: radial velocity of $V_h = 732$~km\,s$^{-1}$, HI line width of 32~km\,s$^{-1}$, HI-flux of 0.70~Jy\,km\,s$^{-1}$, and kinematic NAM distance of 11.0~Mpc. According to the NAM, 43 galaxies fall into the Local Volume with $D_\mathrm{NAM} < 12$~Mpc.  
The average color of the 77 detected dwarf galaxies is
\begin{equation}
    \langle B - V \rangle = 0.57 \pm 0.02 \mathrm{\: mag}.
\end{equation}
while the 24 (out of 28) undetected ones have color of
\begin{equation}
    \langle B - V \rangle = 0.70 \pm 0.04 \mathrm{\: mag}.
\end{equation}
Following \citet{2005AJ....130.2598G}, we determined the hydrogen mass of the galaxy as
\begin{equation}
    M_{\rm HI}/M_{\odot} = 2.36 \times 10^5 D_{\rm Mpc}^2 S_{\rm HI},
\end{equation}
which gives
\begin{equation}
    \log(M_{\rm HI}/M_{\odot}) = 12.33 +2\log D_{\rm Mpc} -0.4 m_{21},
\end{equation}
where the mass is expressed in mass of the Sun and
\begin{equation}
    m_{21}= 17.4 - 2.5 \log(S_{\rm HI})
\end{equation} 
according to the formula adopted in the Third Reference Catalogue of Bright Galaxies~\citep[RC3,][]{1991rc3..book.....D} and in the HyperLEDA database~\citep{mak2014}.
Taking heavy elements into account, the mass of neutral gas in a galaxy is usually represented as  
\begin{equation}
    M_{\rm gas}= 1.4 M_{\rm HI}.
\end{equation}
To estimate the stellar mass, we use the $M/L$ relation
\begin{equation}
    \log(M_*/L_B) = -0.91 + 1.45(B - V),
\end{equation}
justified by \citet{herr2016}, where $B$ and $V$ magnitudes are corrected for the Galactic extinction, 
and the $B$-band luminosity is related to the absolute $B$-magnitude, $M_B$, in the standard way
\begin{equation}
    \log(L_B/L_{\odot})= 0.4 (M_{B,\odot} - M_B).
\end{equation}
Using $M_{B,\odot} = 5.44$~\citep{will2018}, the stellar mass of a galaxy can be expressed as
\begin{equation}
    \log(M_*/M_{\odot}) = 11.27 + 2 \log(D_{\rm Mpc}) + 1.45(B - V) - 0.4B.
\end{equation}

As a result, the equality  $M_{\rm gas} = M_*$ occurs when 
\begin{equation}
    m_{21} = B + 0.96 \pm 0.07 \mathrm{\: mag}.
\end{equation}
Consequently, in the late-type dwarf galaxies with $m_{21} - B < 0\fm96$, the gas component dominates.

According to \citet{2021ApJ...910...69S}, deep observations with the GBT down to a column density of $N_\mathrm{HI} = 6\times10^{17}$~cm$^{-2}$ over a 20~km\,s$^{-1}$ line-width reveal that a fraction of diffuse neutral gas below the $N_\mathrm{HI} < 10^{19}$~cm$^{-2}$ level ranges from 5 to 93\% of the total HI mass. The typical rms noise in our observations is 3.3 mJy, which is at least no worse than the noise level of 3.2--10.4 mJy in the \citet{2021ApJ...910...69S} cubes. Therefore, we can be confident that the total gas masses measured in our survey also include a diffuse component outside the main HI-disk of dwarf galaxies.

The distribution of galaxies in our sample according to $m_{21}$ and $B$ magnitudes is presented in the top panel of Fig.~\ref{figure3}. 
Galaxies undetected in the HI line are shown as open circles with $m_{21}= 20.0$, which corresponds to the upper limit of the flux for these sources of about~0.09~Jy~km~s$^{-1}$. The diagonal lines on Fig.~\ref{figure3} correspond to $\mu = M_{\rm gas}/M_*$ equal to 0.1, 1.0 and 10. 
As can be seen from these data, most detected galaxies are gas-dominated systems with the median gas-to-star mass ratio $\mu = 1.87$.

The hydrogen mass of the detected galaxies lies in the interval $\log(M_{\rm HI}/M_{\odot}) = [5.57$--$9.14]$ with the median of 7.24.  Based on the distribution of undetected galaxies at the upper edge of the diagram, their median $\mu$ value does not exceed 1. Mostly transition-type galaxies (Tr) fall into this category. Note that kinematic distances $D_{\rm NAM}$ were used to estimate the galaxy masses.

The current version of the UNGC catalog contains 275 dwarf galaxies of the late types (T= 9 and 10) with accurate TRGB distance estimates. 
We excluded from consideration three galaxies of the M\,81 group (BK3N, HS\,117, DDO\,82) where the HI line is masked by the local Galactic hydrogen emission, and six galaxies in the Zone of Avoidance with $A_B > 2\fm0$. The distribution of 266 remaining galaxies by $m_{\rm 21}$ and $B-A_B$ magnitudes is presented in the bottom panel of Fig.~\ref{figure3}.  It demonstrates that 97\% of late-type dwarfs reside in the range $\mu  = [0.1$--$10]$ with the median of about $\mu = 1.5$. The most gas-rich dIrr galaxies (And\,IV and ESO\,215-009) have $\mu = 32$ and 26, respectively. The galaxies we detected (upper panel in Fig.~\ref{figure3}) tend to be located in the upper right corner on the bottom panel.

\begin{figure}
    \centering
    \includegraphics[width=\linewidth]{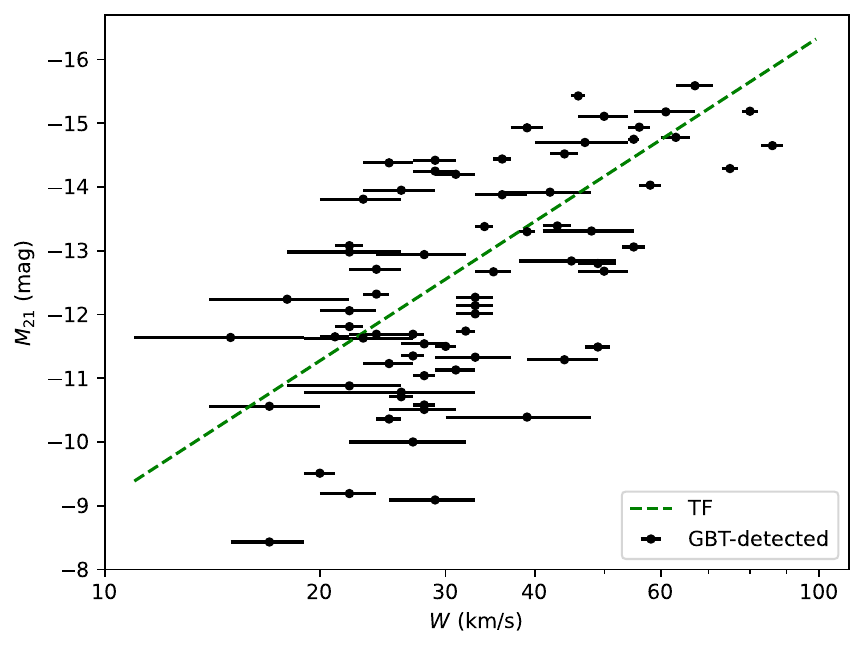}
    \includegraphics[width=\linewidth]{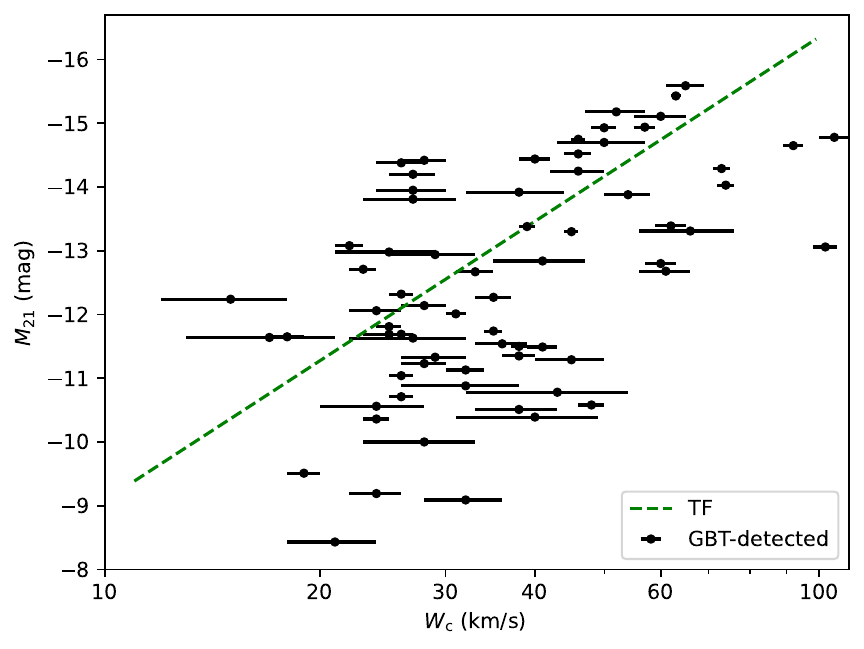}
    \caption{{\it Upper panel.} Magnitude vs. line width relation for nearby dwarf galaxies detected with the GBT. The horizontal bars indicate the standard errors of $W_{50}$. The dashed line  displays the Tully-Fisher relation, eq.(10). {\it Bottom panel.} Magnitude vs. line width relation for the same galaxies but with a correction for galaxy inclination according to eq.(13).} \label{figure4}
\end{figure}

Being mainly gas-dominated systems, late-type dwarfs show the expected correlation between their hydrogen mass and HI line width $W_{50}$, which is the gaseous analogue of the Tully-Fisher relation~\citep{tul1977} between the integral absolute magnitude of galaxy, $M_B$, and $W_{50}$. The upper panel of Fig.~\ref{figure4} displays the relation between  $M_{21}= m_{21}-5 \log(D_{\rm Mpc}) - 25$ and $W_{50}$ for 71 detected galaxies.\footnote{Between ``the absolute HI magnitude''  $M_{21}$ and mass $M_\mathrm{HI}$ there is the equality  $\log(M_{\rm HI})= 2.33 -0.4 M_{21}.$}

Here we excluded the galaxy Dw\,1615+5422 (a peculiar spiral system) and five reddish galaxies (UGC\,4918, PGC\,154449, KUG\,1013+414,  Dw\,1214+2945, and UGCA\,361) that have a smooth structure and low gas content ($M_{gas}/M_* < 0.1$). The horizontal bars indicate the standard errors of $W_{50}$ given in Table~\ref{table1}.  The dashed line in the panel indicates the relation
\begin{equation}\label{cTF}
    M_B = - 7.27 (\log W_{50} - 2.5) -19.99
\end{equation}
according to \citet{tul2008} that has been obtained for luminous disc-dominated galaxies.  As follows from these data, dwarf galaxies satisfactorily follow the classical Tully-Fisher relation if $M_{21}$ magnitudes are used instead of $M_B$.

Fig.~\ref{figure4} allows one to estimate the distances of dwarf irregular galaxies with a typical error of $\sigma(\log D) = 0.27$. 
The scatter of late-type galaxies in Fig.~\ref{figure4} is not a result of measurement errors of $W_{50}$ and $S_{\rm HI}$, but is due to the irregular shape of dwarfs caused by turbulent motions in them. One would hope that by taking into account the inclination angle ($i$) of the galaxy's rotation axis with respect to the line of sight, the dispersion in the $M_{21}$ vs. $W_{50}$ diagram could be reduced. If the galaxy has the shape of an ellipsoid of rotation with a true axial ratio $(b/a)_0$ and an apparent axial ratio $b/a$, then the correction of $W_{50}$ for the inclination is expressed as
\begin{equation}
    W_{50}^c = W_{50}/\sin i,
\end{equation}
where
\begin{equation}
    \sin^2 i = \frac{ 1-(b/a)^2}{1-(b/a)_o^2}. 
\end{equation}
Since the typical axial ratio for irregular dwarfs is $(b/a)_o = 0.6$~\citep{roy2013,kar2017}, the correction for inclination takes the form
\begin{equation}
    W_{50}^c = \frac{0.80 W_{50}}{\sqrt{1-(b/a)^2}}. 
\end{equation}         

The bottom panel of Fig.~\ref{figure4} exhibits the distribution of irregular dwarfs by $M_{21}$ and $W_{50}^c$ after accounting the correction for their inclination angle. Taking into account the inclination does not reduce the dispersion, but instead slightly increases it. The probable cause for the scatter is the difference in the shape of irregular dwarfs from the ellipsoid of rotation, especially in relation to the shape of their gas component~\citep{roy2010}. 

In~\cite{kar2017} the gaseous Tully-Fisher relation have been studied for 206 LV galaxies. The regression line for their distribution over hydrogen mass and HI line width $W_{50}$ has a slope of 2.08 and $\sigma(\log M_\mathrm{HI}) = 0.43$. Passing from $W_{50}$ to $W_{50}^c$ leaves the regression slope and the dispersion nearly the same. In the current study, no regression was performed, instead we used the classical Tully-Fisher relation~(\ref{cTF}) with slope of 2.91 which gives $\sigma(\log M_\mathrm{HI}) = 0.54$ for our data. The difference in the slope is negligible, especially considering that our range of widths is much smaller (by a factor of 5), and that the slope is probably flattened by high-mass galaxies. It can be seen from figure 9 in~\cite{kar2017} that the dispersion increases appreciably while passing from massive spirals to dwarfs especially for galaxies with $W_{50} \lesssim 30$ km s$^{-1}$: there is a noticeable deficit of HI mass in this region compared to the regression line. The majority of the galaxies in our sample are in this region, resulting in a larger dispersion.

Our radial velocity measurements led to the discovery of some dwarf galaxies probably associated with the bright LV spirals: NGC\,628, NGC\,2787, NGC\,3556, NGC\,4490, NGC\,4594, and NGC\,5055. The 4 new supposed satellites around their host galaxies: NGC\,2787, NGC\,3556, NGC\,4490 and NGC\,5055 have a mean linear projected separation of 159 kpc and a mean-square radial velocity difference of 57~km~s$^{-1}$ with respect to their hosts giving the average total-mass-to-K-band luminosity of the spiral galaxies to be $(6.9\pm1.6)$ in the solar units. Also, four dwarf galaxies (Dw\,1234$-$1142, LV\,J1235$-$1104, Dw\,1235$-$1216, and PGC\,974136) are likely new members of the group around the most luminous LV galaxy, NGC\,4594 (Sombrero).

\section{Conclusions}
\label{sec:Conclusion}

Our HI survey of 105 nearby late-type dwarf galaxies using the GBT has produced 77 detections (73~\% of the sample). The median HI-parameters of the detected galaxies are $S_\mathrm{HI}=0.69$~Jy\,km\,s$^{-1}$, $V_\mathrm{HI}=732$~km\,s$^{-1}$, and $W_\mathrm{50} = 32$~km\,s$^{-1}$. About half of them are members of the Local Volume having kinematic distances $D_\mathrm{NAM} < 12$~Mpc and low hydrogen masses in the range of $\log(M_\mathrm{HI}/M_\sun) = [5.57$--$9.14]$.  
Most of the detected dwarfs are gas-dominated objects with the median gas-to-stellar mass ratio of 1.87. In general, the dwarf galaxies follow the classic Tully-Fisher relation, obtained for large disk-dominated spiral galaxies, if their $m_{21}$ magnitudes are used instead of B-magnitudes. This allows one to make rough estimates of the distances of gas-rich dwarfs with a typical error of $\sigma(\log D) \sim 0.27$. 
  
Among the remaining 28 undetected galaxies, most are Tr-type galaxies (transition between dSph and Irr). This category of dwarfs is distinguished by a fairly smooth structure, redder $B-V$ color index, and a weak UV flux. Five of the 28 undetected targets turned out to be very nearby galaxies, whose HI flux is masked by the bright emission of local hydrogen from the Milky Way.

\begin{acknowledgments}

AEN, IDK, DIM, and MIC are supported by the Russian Science Foundation grant \textnumero~24--12--00277.

The National Radio Astronomy Observatory is a facility of the National Science Foundation operated under cooperative agreement by Associated Universities, Inc.

The authors thank Toney Minter for helpful discussions and the GBT operators for ensuring smooth operations during observing sessions. 

This work has made use of the DESI Legacy Imaging Surveys (\url{http://www.legacy/surveys.org/}), the Galaxy Evolution Explorer (GALEX, \url{http://www.galex.caltech.edu/index.html}), 
HyperLEDA database (\url{http://leda.univ-lyon1.fr/}), and the current version of the Local Volume galaxy database (\url{http://www.sao.ru/lv/lvgdb}). 
\end{acknowledgments}

\bibliographystyle{aasjournal}
\bibliography{main}

\end{document}